\documentclass[conf]{new-aiaa}
\usepackage[utf8]{inputenc}

\usepackage{graphicx}
\usepackage{amsmath}
\usepackage[version=4]{mhchem}
\usepackage{siunitx}
\usepackage{pgfplots}
\usepackage{tikz}
\pgfplotsset{compat=newest} 
\usepackage{placeins} 
\usepackage{flafter}  
\usepgfplotslibrary{units} 
\usepackage{longtable,tabularx}
\title{Application of Modal Filtering in Discontinuous Spectral Element Method for Simulation of Compressible Channel Flow at Different Reynolds Numbers}

\author{Ahmad Peyvan,\footnote{Ph.D Candidate, Mechanical and Industrial Engineering Department} Zia Ghiasi,\footnote{Ph.D, Mechanical and Industrial Engineering Department} Dongru Li,\textsuperscript{*} Jonathan Komperda,\textsuperscript{*} Farzad Mashayek\footnote{Professor, Mechanical and Industrial Engineering Department, mashayek@uic.edu}}
\affil{University of Illinois at Chicago, Chicago, IL, 60607, U.S.A.}

\begin{document}

\maketitle
\begin{abstract}
Large eddy simulation of incompressible turbulent flow has been extensively investigated; hence, a variety of models suited for different numerical schemes have been developed. In the case of compressible flow, the modeling is more challenging due to the numerous terms that arise in the filtered Navier-Stokes equations. Recently, modal explicit filtering was implemented in the discontinuous spectral element method (DSEM) to mimic the effects of sub-filtered scales for compressible flows. The method is computationally inexpensive since it is implemented in the DSEM code using a well-established math library. It has been successfully applied to simulations of compressible decaying isotropic turbulence and turbulent channel flow. In this study,  channel flows with three different friction Reynolds numbers are simulated using explicit modal filtering in DSEM to assess the performance of the new method by comparing turbulent statistics and friction Reynolds numbers to those from DNS. The model predicts the friction Reynolds number with the maximum of $\textrm{2.16\%}$ error with respect to DNS. Although the model generates promising results for the average velocity profile compared to DNS, it requires adjustment of filter strength and frequency to predict rms statistics accurately.
\end{abstract}

\section{Nomenclature}

{\renewcommand\arraystretch{1.0}
\noindent\begin{longtable*}{@{}l @{\quad=\quad} l@{}}
$u_i$  & velocity components \\
$\rho$ & density \\
$x_j$& Cartesian coordinates \\
$X_j$& mapped coordinates \\
$t$ & time \\
$p$ & pressure \\
$P$ & polynomial order \\
$\delta_{ij}$   & Kronecker delta \\
$\delta$   & channel half-height\\
$\nu$ & kinematic viscosity \\
$e$ & total energy \\
$q_j$ & heat flux vector \\
$\tau_{ij}$ & viscous stress tensor \\
$Re_f$   & reference Reynolds number \\
$Re_\tau$   & friction Reynolds number \\
$L_x$   & stream-wise domain length  \\
$L_y$   & wall-normal domain length  \\
$L_z$   & span-wise domain length  \\
$Pr$  & Prandtl number \\
$M_f$  & reference Mach number \\
$\gamma$  & heat capacity ratio \\
$T$  & non-dimensional temperature
\end{longtable*}}

\section{Introduction}

The growing power of computers has made the direct numerical simulation (DNS) of fundamental turbulent flows\cite{li2019compressibility,abtahi2019porous}, such as decaying isotropic turbulence\cite{ghiasi2019modal}, shear layers\cite{li2019compressibility}, and plane channel flows\cite{ghiasi2019modal}, achievable. However, performing DNS of complex-geometry flows and capturing eddy motions of all scales still remains infeasible. Therefore, large eddy and Reynolds average simulations\cite{peyvan2016axial} became the center of attention for the recent decades. In LES, the motions of large eddies are resolved, whereas the small scale effects are modeled \cite{pope2001turbulent}. In the traditional LES approach,  a low-pass spatial filter is applied to the Navier-Stokes equations to divide the flow structure to resolved and subgrid scales (SGS). The commutative filtering operation adds SGS terms to the Navier-Stokes equation. In the simplest form, the SGS terms are calculated using eddy viscosity models such as Smagorinsky model \cite{smagorinsky1963general}. This type of filtering is called implicit filtering since no explicit filtering is performed and the coarse grid length is the filter cutoff length scale.

The eddy viscosity models assume that the Reynolds stress tensor is proportional to the strain tensor, whereas Bardina et al. \cite{bardina1980improved} showed the principal axes of strain and Reynolds stress tensor are not aligned. Therefore, they proposed a new SGS model, called scale similarity, which uses two levels of filtering to calculate the SGS term. The coarse mesh itself applies the first level of filtering with a cutoff length equal to the grid size. The second level of filtering is performed using an explicit filter on the resolved solution to calculate the larger scales with a greater cutoff length. Scale similarity does not ensure a net positive rate of energy transfer from large scales to small scales, and it does not dissipate energy \cite{bardina1980improved}. As a result, eddy viscosity is combined with scale similarity  to obtain a high correlation between the exact and modeled Reynolds stress tensor, as well as to model the energy dissipation. Germano et al. \cite{germano1991dynamic} also applied two-level filtering to determine the Smagorinsky constant dynamically. The only input parameter of Germano's model is the ratio of filter cutoff lengths, which can be variable for different flows. Generally, whenever two levels of filtering is needed, explicit filtering proves to be useful.

Large eddy simulation of incompressible flow has been studied extensively using SGS modeling. For compressible flows, a few SGS models have been developed, because of the complexity of SGS terms in the filtered energy equation. Moin et al. \cite{moin1991dynamic} extended Germano's dynamic SGS model to LES of compressible flows and transport of a scalar. They used DNS data of isotropic turbulence, homogeneous shear flow, and turbulent channel flow to evaluate the SGS turbulent Prandtl number. Erlebacher et al. \cite{erlebacher1992toward} developed a compressible version of mixed scale-similarity in terms of Favre-filtered fields, to model subgrid-scale tensor for compressible decaying isotropic turbulence. They obtained the model by using a linear combination of the Smagorinsky and scale-similarity models, to calculate the Reynolds stress tensor and heat flux vector, respectively. Lenormand et al. \cite{lenormand2000large} used two subgrid-scale models, a compressible extension of Smagorinsky model and Bardina-selective mixed scale model, to simulate subsonic and supersonic channel flow. Their model showed a good agreement with the experiment and DNS results. Martin et al. \cite{martin2000subgrid} applied several mixed and eddy-viscosity models for the momentum and energy equation using the decaying isotropic turbulence DNS data. They assessed the performance of LES models to predict closure terms in internal energy, enthalpy, and total energy forms of the energy equation and found that mixed similarity models perform better than eddy-viscosity models. Lodato et al. \cite{lodato2013discrete} applied a class of constrained discrete filter operators to calculate mixed scale similarity model terms for LES of channel flow. They defined the filter operator based on the work of Vasilyev et al. \cite{vasilyev1998general} to minimize the commutation error for high-order discretization technique with non-uniform solution point distribution. They investigated the performance of the SGS model calculated using the constrained filter in fully developed channel flow and obtained a good result for average and rms statistics compared to DNS. Calculation of SGS terms in LES could add a significant computational overhead to simulation of compressible flows in complex geometries.

Recently, LES of turbulent flows without SGS modeling has gained more attention for compressible flows since it does not add complex SGS terms to the governing equations, and does not increase computational cost. Grinstein and Fureby \cite{grinstein2002recent} used a new class of LES, known as monotonically integrated LES  (MILES), for several cases of fully developed channel flows. In this approach, the numerical dissipation plays the role of SGS terms and dissipates the turbulent energy. This method is not applicable to high-order numerical schemes that have little dissipation and small margins of stability, such as DSEM. As an alternative for  numerical dissipation, one can employ an explicit low-pass filter to drain the energy of small scales in high order numerical schemes. In this context, Gassner and Beck \cite{gassner2013accuracy} investigated the accuracy of a high-order discretization scheme, discontinuous Galerkin spectral element (DGSEM), for under-resolved simulation of the Taylor-Green vortex problem. They used a simple exponential-based modal filter to eliminate the energy at high wavenumbers, consequently, reducing aliasing error. However, they did not investigate the performance of their filtering method for wall-bounded flows.  Flad et al. \cite{flad2016simulation} employed a cell local projection filter to stabilize the simulation of under-resolved decaying isotropic turbulence and transitional Taylor-Green vortex using DGSEM. The stabilization technique is performed by projection of the solution vector from polynomial degree M to a lower degree N. Also, Winters et al. \cite{winters2018comparative} focused on two methods of stabilizing an under-resolved turbulence simulation of inviscid Taylor-Green vortex flow using DGSEM. The first strategy was over-integration, where the quadrature accuracy is improved by increasing the number of quadrature points. The second approach is built upon the concept of the split form of the advective terms in the governing equations. In the split form, the nonlinear terms of Navier-Stokes equation are derived as the average of conservative and non-conservative variables. They showed that the split form could prevent the accumulation of energy at high-wavenumbers. The split form and over integration methods require additional computational effort, which is a disadvantage in LES of complex geometry flows. 

Ghiasi et al. \cite{ghiasi2019modal} implemented the modal filtering method in a discontinuous spectral element scheme and tested its performance by simulating decaying isotropic turbulence and turbulent channel flow with a specific friction Reynolds number. The DSEM scheme has been employed for simulating  various flow configuration such as reacting and non-reacting Taylor-Green vortex flow \cite{komperda2020hybrid}, spherical explosions \cite{komperda2020filtered}, channel flow\cite{ghiasi2019modal}, backward facing step \cite{ghiasi2018near}, and isotropic decaying turbulence\cite{ghiasi2019modal}. In this work, we investigate the application of the modal filtering method with no SGS model to assess the performance of the model for LES of channel flow with various Reynolds numbers. We apply the modal filter locally on each element, resulting in a low computational cost and ease of implementation. The explicit modal filtering extends the application of DSEM to LES of complex geometry flows.   

\section{Methodology}

\subsection{Governing Equations}
In this paper, we solve full compressible Navier-Stokes equations in a conservative form. We solve the non-dimensional form, which is presented with Cartesian tensor notation as

\begin{equation}
\label{eqation1}
\frac{\partial\rho}{\partial t}+\frac{\partial( \rho u_j)}{\partial x_j}=0,
\end{equation}
\begin{equation}
\label{eqation2}
\frac{\partial (\rho u_i)}{\partial t}+\frac{\partial (\rho u_i u_j+p\delta_{ij})}{\partial x_j}=\frac{\partial \tau_{ij}}{\partial x_j},
\end{equation}
\begin{equation}
\label{eqation3}
\frac{\partial (\rho e) }{\partial t}+\frac{\partial( (\rho u_j e+p)u_j)}{\partial x_j}=-\frac{\partial q_j}{\partial x_j}+\frac{(\tau_{ij}u_i)}{\partial x_j}.
\end{equation}
The total energy ($e$), viscous stress tensor ($\tau_{ij}$), and heat flux vector ($q_j$) are expressed, respectively, as 

\begin{equation}
\label{eqation4}
\rho e= \frac{p}{\gamma -1}+\frac{1}{2}\rho u_k u_k,
\end{equation}
\begin{equation}
\label{eqation5}
\tau_{ij}=\frac{1}{Re_f}\Big(\frac{\partial u_i}{\partial x_j}+\frac{\partial u_j}{\partial x_i}-\frac{2}{3}\frac{\partial u_k}{\partial x_k}\delta_{ij}\Big),
\end{equation}
\begin{equation}
\label{eqation6}
q_j= \frac{1}{(\gamma -1)Re_f Pr M_f^2}\frac{\partial T}{\partial x_j}.
\end{equation}
Here,  $Re_f=\rho_f^*U_f^*L_f^*/\mu_f^*$ is reference Reynolds number, which is calculated based on reference density, $\rho_f^*$, reference velocity, $U_f^*$, reference length, $L_f^*$, and reference dynamic viscosity, $\mu_f^*$. The Prandtl number is defined as $Pr=\mu_f^*C_p^*/k^*$, where $C_p^*$ and $k^*$ are the constant-pressure specific heat capacity and thermal conductivity, respectively. The superscript $*$ denotes dimensional quantities. We use a non-dimensional form of the equation of state 

\begin{equation}
\label{eqation7}
p=\frac{\rho T}{\gamma M_f^2},
\end{equation}
to close the system of equations. The reference Mach number, $M_f=U_f^*/c_f^*$, is defined with $c_f^*=\sqrt[]{\gamma R T_f^*}$ as reference speed of sound, where $R$ and $T_f^*$ are the gas constant and reference temperature, respectively. We can express the Navier-Stokes equations in a vector form as

\begin{equation}
\label{eqation8}
\frac{\partial \vec{Q}}{\partial t}+ \frac{\partial \vec{F}_i^a}{\partial x_j}=\frac{\partial \vec{F}_i^v}{\partial x_j},
\end{equation}
where

\begin{equation}
\label{eqation9}
\vec{Q}=\begin{pmatrix}
\rho \\ \rho u_1 \\ \rho u_2 \\ \rho u_3 \\ \rho e
\end{pmatrix}, \quad \vec{F}_i^a=\begin{pmatrix}
\rho u_i \\ p\delta_{i1}+\rho u_1 u_i \\ p\delta_{i2}+\rho u_2 u_i \\ p\delta_{31}+\rho u_3 u_i \\ u_i(\rho e+p)
\end{pmatrix}, \quad \vec{F}_i^v=\begin{pmatrix}
0 \\ \tau_{i1} \\ \tau_{i2} \\ \tau_{i3} \\ -q_i+u_k\tau_{ik}
\end{pmatrix}.
\end{equation}
In Eqs.~\eqref{eqation8} and ~\eqref{eqation9}, $\vec{Q}$ is the solution vector, and $\vec{F}_i^a$ and $\vec{F_i^v}$ are advective and viscous flux vectors, respectively.

\subsection{Numerical Method}
In this work, we employ the discontinuous spectral element method to solve Eqs.~\eqref{eqation7}and~\eqref{eqation8} numerically. In DSEM, first, the physical domain is divided into non-overlapping elements. Then each element is mapped onto a unit cube with $[0,1]^3$ dimensions using isoparametric mapping. Equation ~\eqref{eqation8}, in the mapped space, is

\begin{equation}
\label{eqation10}
\frac{\partial \tilde{Q}}{\partial t}+ \frac{\partial \tilde{F}_i^a}{\partial X_j}=\frac{\partial \tilde{F}_i^v}{\partial X_j},
\end{equation}
where
\begin{equation}
\label{eqation11}
\tilde{Q}=J\vec{Q}, \quad \tilde{F}_i^a=\frac{\partial X_i}{\partial x_j}\vec{F}_j^a, \quad \tilde{F}_i^v=\frac{\partial X_i}{\partial x_j}\vec{F}_j^v.
\end{equation}
In Eqs.~\eqref{eqation10} and ~\eqref{eqation11}, tilde indicates a mapped vector, and $J$ is the Jacobian of the mapping. The term $\frac{\partial X_i}{\partial x_j}$ is the transformation metric, and $X_i$ and $x_j$ are mapped and physical space coordinates, respectively. In each element, Eq. \eqref{eqation10} is discretized on a staggered Chebyshev grid \cite{kopriva1996conservative}. In DSEM, for one-dimensional grid, we calculate the solution and flux values on a distribution of Gauss and Gauss-Lobatto collocation points, respectively. Gauss and Gauss-Lobatto distributions in the mapped space on interval $[0,1]$ are represented  as
\begin{equation}
\label{eqation12}
X_{i+\frac{1}{2}}=\frac{1}{2}\Big[1-\cos\Big(\frac{i+1}{2(P+1)}\pi\Big)\Big], \quad i=0, \ldots,P,
\end{equation}
\begin{equation}
\label{eqation13}
X_{i}=\frac{1}{2}\Big[1-\cos\Big(\frac{i}{P+1}\pi\Big)\Big], \quad i=0, \ldots,P+1,
\end{equation}
In Eqs. \eqref{eqation12} and \eqref{eqation13}, $P$ is the approximation polynomial order. The solution ($\tilde{Q}$) and fluxes ($\tilde{F}$) are evaluated with a high-order Lagrange polynomial basis on each element 
\begin{equation}
\label{eqation14}
\tilde{Q}(X_1,X_2,X_3)=\sum_{i=0}^{P}\sum_{j=0}^{P}\sum_{k=0}^{P}\tilde{Q}_{i+1/2,j+1/2,k+1/2}h_{i+1/2}(X_1)h_{j+1/2}(X_2)h_{k+1/2}(X_3),
\end{equation}
\begin{equation}
\label{eqation15}
\tilde{F}(X_1,X_2,X_3)=\sum_{i=0}^{P}\sum_{j=0}^{P}\sum_{k=0}^{P}\tilde{F}_{i,j,k}h_{i}(X_1)h_{j}(X_2)h_{k}(X_3).
\end{equation}
The terms $h_{i+1/2}$ and $h_i$ are Lagrange interpolation polynomials for Gauss and Gauss-Lobatto points, respectively. The advective fluxes are patched on the element interfaces using an approximate Roe's Reimann solver \cite{toro2013riemann}, while the viscous fluxes are patched using an arithmetic mean from both sides of the interface values \cite{jacobs2003numerical}. In the DSEM, the solution is updated in time using a fourth-order, low storage Runge-Kutta scheme \cite{jacobs2003numerical} after the fluxes are patched.

\subsection{Filtering Procedure}
In DSEM, the solution is represented with a summation of Lagrange polynomials, which are locally constructed on Gauss points, Eq. \eqref{eqation14}, in the nodal mapped space. The solution function can also be represented in modal space using orthogonal basis such as trigonometric functions. Inside a 1D element, a local solution function, $q(X,t)$, at a specific time, with a polynomial order $P$ can be expressed in nodal form as
\begin{equation}
\label{eqation16}
q(X,t)=\sum_{i=0}^{P}\tilde{q}_{i+1/2}(t)h_{i+1/2}(X),
\end{equation}
where $\tilde{q}_{i+1/2}(t)$ are the solution values on Gauss points at  time $t$ and $h_{i+1/2}(X)$ is the Lagrange polynomial of order $P$. The modal expansion of the solution function is defined as
\begin{equation}
\label{eqation17}
q(X,t)=\sum_{l=0}^{P}\hat{q}_l(t)\phi_l(X),
\end{equation}
where $\phi_l(X)$ is the orthogonal basis function, and $\hat{q}_l(t)$ is the expansion coefficient. In the set $\{\hat{q_l}\}_{l=0}^{l=P}$, each successive function represents a higher mode with a higher spatial frequency. We substitute Eq. \eqref{eqation16} into Eq. \eqref{eqation17} and evaluate the expression at the mapped coordinates of the Gauss points, $X=X_{i+1/2}$, to derive
\begin{equation}
\label{eqation18}
\tilde{q}(X_{i+1/2},t)=\frac{1}{2}\hat{q}_0(t)+\sum_{l=1}^{P}\hat{q}_l(t)\cos\Big[\frac{l\pi}{(P+1)}\big(i+\frac{1}{2}\big)\Big], \;\; i= 0,...,P.
\end{equation}
Equation \eqref{eqation18} is called the inverse discrete Chebyshev transform (iDChT) \cite{kopriva2009implementing}. The discrete Chebyshev transform (DChT) \cite{kopriva2009implementing} is defined as 
\begin{equation}
\label{eqation19}
\hat{q}_l(t)=\frac{2}{(P+1)}\sum_{i=0}^P\tilde{q}(X_{i+1/2},t)\cos\Big[\frac{l\pi}{(P+1)}\big(i+\frac{1}{2}\big)\Big], \quad i= 0,...,P.
\end{equation}
Equation \eqref{eqation19} transforms nodal solution values, $\tilde{q}(X_{i+1/2},t)$, to modal values, $\hat{q}_l(t)$. The modal expression of the solution function in 3D space is simply the tensor-product of the 1D basis functions
\begin{equation}
\label{eqation20}
\tilde{q}(X_1,X_2,X_3,t)=\sum_{k=0}^P\sum_{l=0}^P\sum_{m=0}^P\hat{q}_{klm}(t)\phi_{k}\phi_{l}\phi_{m}.
\end{equation}
In this representation, $\hat{q}_{klm}(t)$ is the modal expansion coefficient tensor that expresses modes, the so-called modal tensor. Modal explicit filtering works in three steps. First, the mapped nodal solution is transformed to modal space using the DChT. Second, we  set the highest frequency modes, i.e. some components of the modal tensor, to zero to drain energy from the turbulent domain. Third, we transform the modal solution back to the nodal space by applying iDChT. Here, we apply the filtering based on a parameter, $P_f$, which indicates the strength of filtering, and defines the filtering process as
\begin{equation}
\label{eqation21}
\hat{q}(t)_{klm}=0 \quad \forall \;\;\; \{k,l,m\}, \quad \textrm{where} \quad \max\{k,l,m\}>P-P_f.
\end{equation}
Considering an example of $P_f=1$ in a simulation with $P=6$, a total number of $3P^2+3P+1=127$ components of the modal tensor will be removed.

\section{Channel Flow}
In this section, we present simulations of fully developed turbulent channel flows to investigate the performance of the new LES approach, i.e., modal filtering, with wall-bounded flows. First, we explain the problem configuration. Then we show the simulation results, including Reynolds averaged velocity profiles and Reynolds mean stress components, and compare DNS results. 
\subsection{Problem Setup}
\begin{figure}[hbt!]
\centering
\includegraphics[width=.5\textwidth]{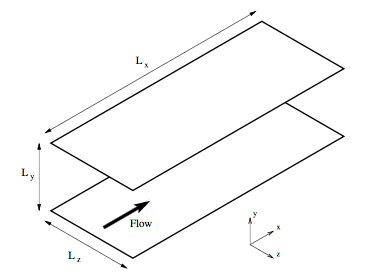}
\caption{Schematics of physical domain of channel flow}
\label{fig1}
\end{figure}

A schematic of the physical domain of the channel flow is shown in Fig.~\ref{fig1}. Periodic boundary conditions are defined at the boundaries of the domain in the stream-wise and span-wise directions. A no-slip isothermal wall boundary condition is applied at boundaries in the wall-normal direction. We also employ a time-dependent forcing term, introduced by Lenormand et al.\cite{lenormand2000large}, to maintain a constant mass flow rate. The DNS of Ghiasi et al. \cite{ghiasi2018near}, Moser et al.\cite{moser1999direct}, and Lee and Moser \cite{lee2015direct} are used as reference for $Re_\tau=204$, $395$, and $544$, respectively. In all the simulations, the bulk velocity, $\bar{U}$, the bulk density, $\bar{\rho}$, the wall temperature, $T_w$, and the channel half-height, $\delta$, are chosen as the reference values. The friction Reynolds number is defined as, $Re_\tau=u_\tau\delta/\nu$, where $\nu$ is kinematic viscosity and 
\begin{equation}
\label{eqation23}
u_\tau=\sqrt[]{\frac{\nu \frac{\partial u}{\partial y}\big|_{wall}}{Re_f}}
\end{equation}
is the friction velocity. The reference Mach number, $M_f$, which is calculated based on the reference temperature and velocity, for CHN200, CHN395, and CHN540 is $0.4$, $0.3$, and $0.4$, respectively. The heat capacity ratio is constant, $\gamma=1.4$, and the Prandtl number is $Pr=0.72$. The size of the domain for each friction Reynolds number is selected such that it encompasses the largest scales. Table 1. presents the computational domain size and grid information for each case. 

\begin{table}[hbt!]
\caption{\label{tab:table1} Domain size and grid configuration of channel flow cases}
\centering
\begin{tabular}{lcccccccccc}
\hline
Case& $Re_f$&$L_x\times L_z$& $N_x\times N_y\times N_z$ & \shortstack{Polynomial \\ order }& \shortstack{Total solution \\ points}& $y_{min}^+$&$\overline{\Delta x^+}$& $\overline{\Delta z^+}$&\shortstack{Points \\in $y_{10}^+$} &$\lambda$ \\\hline
CHN200&3,266& $5.61\delta\times2\delta$& $4\times6\times4$& 6& 32,928&0.255&41&15& 3&4.0\\
CHN395&7,095& $2\pi\delta\times\pi\delta$& $10\times12\times15$& 6& 617,400& 0.169& 35&12&5&4.0 \\
CHN540&10,000& $4\pi\delta\times2\pi\delta$ & $20\times12\times30$& 6& 2,469,600& 0.167&49&16& 5&4.5\\
\hline
\end{tabular}
\end{table}
In Table.~\ref{tab:table1}, $L_x$ and $L_z$ are normalized by the channel half-height, $\delta$. The terms $N_x$, $N_y$, and $N_z$ indicate the number of elements in each direction. We denote simulations of $Re_\tau=204$, $Re_\tau=395$, and $Re_\tau=544$ with CHN200, CHN395, and CHN540, respectively. Since we employ a non-uniform solution point distribution in an element, we define an average grid size normalized by wall scales in the $x$ and $z$ directions as $\overline{\Delta x^+}=\frac{u_\tau L_x}{\nu(N_x(P+1))}$ and $\overline{\Delta z^+}=\frac{u_\tau L_z}{\nu(N_z(P+1))}$. The superscript ($^+$) means the value is scaled with friction velocity, $u_\tau$. In the term $y^+_{min}=\frac{u_\tau y_{min}}{\nu}$, $y_{min}$ is the wall-normal distance of the first solution point near the wall. The grid resolution for all the cases falls within the suitable range required for LES without wall modeling proposed by Choi and Moin \cite{choi2012grid}. Finally, $\lambda$ is a constant for determining element clustering ratio near the wall in the hyperbolic function  
\begin{equation}
\label{eqation22}
\frac{y_m}{L_y}=\frac{1}{2}\Bigg(1-\frac{\tanh\big[\lambda\big(\frac{1}{2}-\frac{m}{N_y}\big)\big]}{\tanh\big[\frac{\lambda}{2}\big]}\Bigg),\;\;\; m=0,...,N_y,
\end{equation}
where $y_m$ is the wall-normal coordinate of an element interface.

\subsection{Results}
We performed a series of LES of periodic channel flow for three friction Reynolds numbers to investigate the performance of modal filtering method by comparing the averaged and the root-mean-square statistics with the DNS results. For each case, a coarse DNS simulation is performed and then a modal filter with strength $P_f=1$ is applied every 100-time steps to determine the effect of filtering on the solution. Giasi et al. \cite{ghiasia2017modal} assessed the effect of the strength ($P_f$) and the frequency of filtering on the predicted friction Reynolds number. For the filter strength effect, they simulated a channel flow with $P=6$, but different mesh resolution, and concluded that $P_f=1$ produces the closest results to DNS regardless of the element size. In this work, the strength of the isotropic filter for all the LES cases is $P_f=1$, meaning the highest modes in all three spatial directions are removed.


In Table~\ref{tab:table2}, we compare the friction Reynolds number calculated using coarse DNS and modal filtering LES with the DNS value. The friction Reynolds number is calculated based on friction velocity, which is directly proportional to the stream-wise velocity profile slope at the wall. We observe that the modal filtering approach predicts the slope with an acceptable error, defined as $\textrm{\%Error}= \frac{Re_\tau(P_f=1 \textrm{ or Coarse DNS})-Re_\tau(\textrm{DNS})}{Re_\tau(\textrm{DNS})}\times 100 $, compared to DNS, given the fact that we used a coarse grid near the wall. Therefore, the modal filtering LES does not require  wall modeling. Moreover, removing the highest modes reduces the skin friction, which leads to reduction of friction Reynolds number compared to coarse DNS.

\begin{table}[hbt!]
\caption{\label{tab:table2} Modal filtering effect on friction Reynolds number}
\centering
\begin{tabular}{lccccc}
\hline
Case& $Re_\tau$ (DNS) & $Re_\tau$ (Coarse DNS)&  $Re_\tau$ ($P_f=1$)&  $\%$Error (Coarse DNS) & $\%$Error ($P_f=1$) \\\hline
CHN200&206.09 & 221.45 &210.35 & 7.45&  2.07\\
CHN395&392.24& 414.32 &400.72  & 5.63&  2.16\\
CHN540& 543.50& 575.49 &545.40 & 5.86&  0.35\\
\hline
\end{tabular}
\end{table}

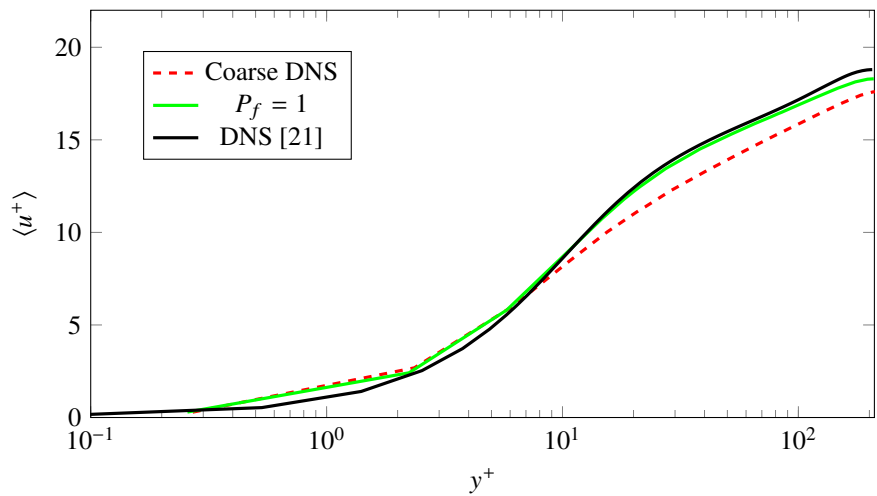
\begin{figure}[hbt!]
  \begin{center}
    \begin{tikzpicture}
  
      \begin{semilogxaxis}[
          height=7cm,width=12cm,
          xlabel=$y^+$,ylabel style={font=}, 
          ylabel=$ \left\langle u^+\right>$,xlabel style={font=},
          legend style={at={(0.2,0.9)},anchor=north}, 
          x tick label style={anchor=north}, 
          xmin=0.1,xmax=210,ymin=0,ymax=22
        ]
        \addplot [dashed,color=red,very thick]
        table[x=y_c,y=u_c,col sep=comma] {Re204ave.csv}; 
        \addplot [color=green,very thick]
        table[x=y_l,y=u_l,col sep=comma] {Re204ave.csv}; 
         \addplot [color=black,very thick]
        table[x=y_d,y=u_d,col sep=comma] {Re204ave.csv}; 
        \legend{Coarse DNS,$P_f=1$,DNS \cite{ghiasi2018near}}
      \end{semilogxaxis}

    \end{tikzpicture}
    \caption{Averaged velocity profile scaled with friction velocity comparison with coarse DNS and DNS of Ghiasi et al. \cite{ghiasi2018near}, for CHN200 }
    \label{fig2}
  \end{center}
  
\end{figure}

In Fig.~\ref{fig2}, the average velocity profile scaled with $u_\tau$ is plotted versus dimensionless wall distance, $y^+$, for the case CHN200. The mean profiles calculated with coarse DNS and $P_f=1$ are compared with the DNS results. In the near wall region, $y^+<8$, a small difference can be observed between $P_f=1$ and the coarse DNS curves; however, the $P_f=1$ curve remains closer to DNS curve. Also, as shown in Table~\ref{tab:table1}, the modal filter calculates $Re_\tau$ more accurately, which means the slope of the mean velocity profile near the wall is improved by the filter. There are three solution points within $y^+=10$, which is substantially less than the 11 points in the DNS simulation. Yet, the general trend of the DNS curve in the viscous sublayer, $y^+<5$, is captured by the modal filtering technique. In the buffer layer, $5<y^+<30$, which is a transition region between the viscosity-dominated and turbulent-dominated region, the $P_f=1$ curves matches the DNS completely. However, in the log-law ($y^+>30,y/\delta<0.3$) and outer layer ($y^+>50$) regions, the modal filtering model slightly under-predicts the velocity profile.

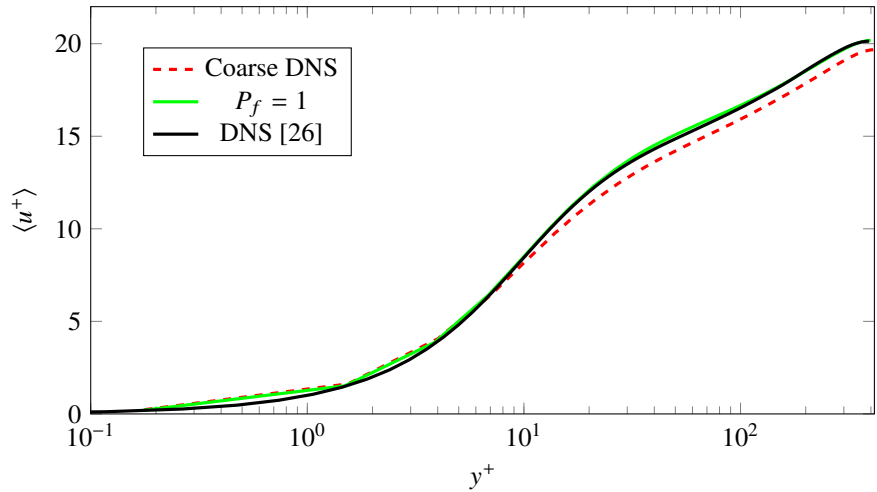
\begin{figure}[hbt!]
  \begin{center}
    \begin{tikzpicture}
  
      \begin{semilogxaxis}[
          height=7cm,width=12cm,
          xlabel=$y^+$,ylabel style={font=}, 
          ylabel=$ \left\langle u^+\right>$,xlabel style={font=},
          legend style={at={(0.2,0.9)},anchor=north}, 
          x tick label style={anchor=north}, 
          xmin=0.1,xmax=415,ymin=0,ymax=22
        ]
        \addplot [dashed,color=red,very thick]
        table[x=y_c,y=u_c,col sep=comma] {Re395ave.csv}; 
        \addplot [color=green,very thick]
        table[x=y_l,y=u_l,col sep=comma] {Re395ave.csv}; 
         \addplot [color=black,very thick]
        table[x=y_d,y=u_d,col sep=comma] {Re395ave.csv}; 
        \legend{Coarse DNS,$P_f=1$,DNS \cite{moser1999direct}}
      \end{semilogxaxis}

    \end{tikzpicture}
    \caption{Averaged velocity profile scaled with friction velocity comparison with coarse DNS and DNS of Moser et al. \cite{moser1999direct}, for CHN395 }
    \label{fig3}
  \end{center}
  
\end{figure}
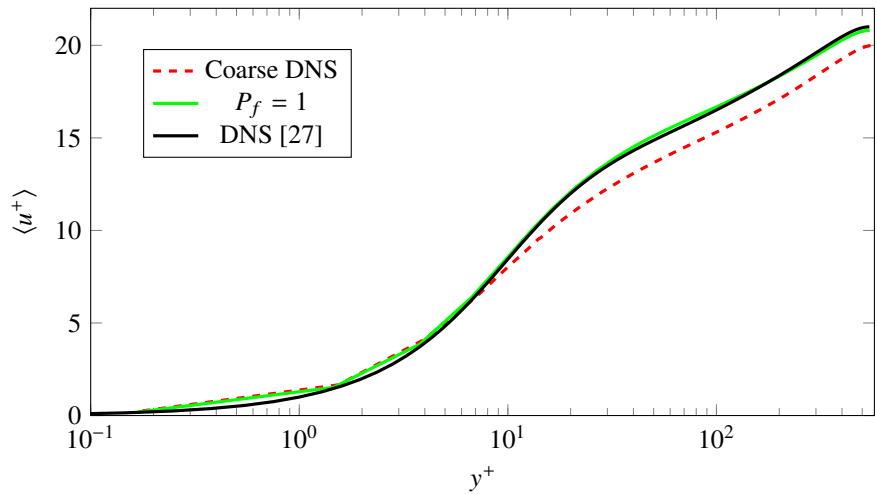
\begin{figure}[hbt!]
  \begin{center}
    \begin{tikzpicture}
  
      \begin{semilogxaxis}[
          height=7cm,width=12cm,
          xlabel=$y^+$,ylabel style={font=}, 
          ylabel=$ \left\langle u^+\right>$,xlabel style={font=},
          legend style={at={(0.2,0.9)},anchor=north}, 
          x tick label style={anchor=north}, 
          xmin=0.1,xmax=572,ymin=0,ymax=22
        ]
        \addplot [dashed,color=red,very thick]
        table[x=y_c,y=u_c,col sep=comma] {Re540ave.csv}; 
        \addplot [color=green,very thick]
        table[x=y_l,y=u_l,col sep=comma] {Re540ave.csv}; 
         \addplot [color=black,very thick]
        table[x=y_d,y=u_d,col sep=comma] {Re540ave.csv}; 
        \legend{Coarse DNS,$P_f=1$,DNS \cite{lee2015direct}}
      \end{semilogxaxis}

    \end{tikzpicture}
    \caption{Averaged velocity profile scaled with friction velocity comparison with coarse DNS and DNS of lee and Moser \cite{lee2015direct}, for CHN540}
    \label{fig4}
  \end{center}
  
\end{figure}

Figure~\ref{fig3} shows the scaled mean velocity profile for case CHN395. The coarse DNS curve is closer to the DNS curve in the CHN395 case compared to the CHN200 since a finer scaled grid size ($\overline{\Delta x^+}$ and $\overline{\Delta z^+}$) is used in CHN395. The modal filter with $P_f=1$ predicts the correct velocity profile and performs better than CHN200. A would be expected, the modal filtering model performance improves by increasing the grid resolution. The $P_f=1$ curve matches with the DNS curve almost everywhere except in the log-law region where it slightly over-predicts the mean profile. Figure~\ref{fig4} presents the average normalized velocity profile for the CHN540 case. The number of points in $y^+=10$ for CHN540 is the same as the case CHN395, but in the span-wise and stream-wise direction, the grid is coarser regarding the normalized grid spacing. In this case, also the modal filtering model predicts the mean profile accurately. 

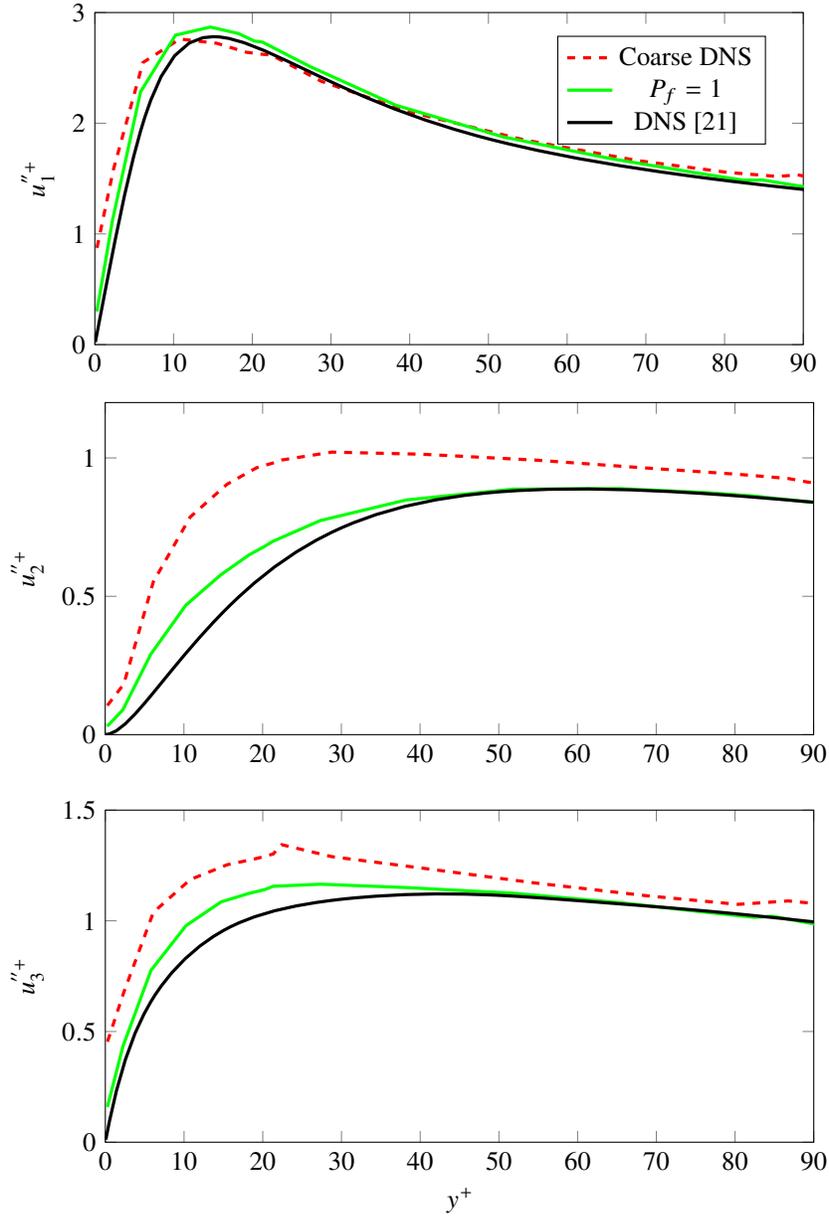
\begin{figure}[hbt!]
  \begin{center}
    \begin{tikzpicture}
  
      \begin{axis}[
          height=6cm,width=11cm,
          ylabel=$u_1^{''+}$,ylabel style={font=},
          legend style={at={(0.8,0.93)},anchor=north}, 
          x tick label style={anchor=north}, 
          xmin=0,xmax=90,ymin=0,ymax=3
        ]
        \addplot [dashed,color=red,very thick]
        table[x=y_c,y=u_c,col sep=comma] {Re204rms.csv}; 
        \addplot [color=green,very thick]
        table[x=y_l,y=u_l,col sep=comma] {Re204rms.csv}; 
         \addplot [color=black,very thick]
        table[x=y_d,y=u_d,col sep=comma] {Re204rms.csv}; 
        \legend{Coarse DNS,$P_f=1$,DNS \cite{ghiasi2018near}}
      \end{axis}

    \end{tikzpicture}
    \\
    \begin{tikzpicture}
  
      \begin{axis}[
          height=6cm,width=11cm,
          ylabel=$u_2^{''+}$,ylabel style={font=},
          legend style={at={(0.8,0.9)},anchor=north}, 
          x tick label style={anchor=north}, 
          xmin=0,xmax=90,ymin=0,ymax=1.2
        ]
        \addplot [dashed,color=red,very thick]
        table[x=y_c,y=v_c,col sep=comma] {Re204rms.csv}; 
        \addplot [color=green,very thick]
        table[x=y_l,y=v_l,col sep=comma] {Re204rms.csv}; 
         \addplot [color=black,very thick]
        table[x=y_d,y=v_d,col sep=comma] {Re204rms.csv}; 
      \end{axis}

    \end{tikzpicture}
    \\
    \begin{tikzpicture}
  
      \begin{axis}[
          height=6cm,width=11cm,
          xlabel=$y^+$,ylabel style={font=}, 
          ylabel=$u_3^{''+}$,xlabel style={font=},
          legend style={at={(0.7,0.9)},anchor=north}, 
          x tick label style={anchor=north}, 
          xmin=0,xmax=90,ymin=0,ymax=1.5
        ]
        \addplot [dashed,color=red,very thick]
        table[x=y_c,y=w_c,col sep=comma] {Re204rms.csv}; 
        \addplot [color=green,very thick]
        table[x=y_l,y=w_l,col sep=comma] {Re204rms.csv}; 
         \addplot [color=black,very thick]
        table[x=y_d,y=w_d,col sep=comma] {Re204rms.csv}; 
      \end{axis}

    \end{tikzpicture}
    \caption{Stream-wise, wall-normal, and span-wise rms velocity fluctuations, compared with DNS of Ghiasi et al.  \cite{ghiasi2018near}, for CHN200}
    \label{fig5}
  \end{center}
  
\end{figure}

The rms fluctuation velocity in $x_i$-direction is defined as $u_i^{''+}=\sqrt[]{\{u_i^{''}u_i^{''}\}/u_\tau^2}$ in which, $\{\}$ denotes Favre-average. The rms velocities determined with modal filtering and coarse DNS are compared with DNS results for $Re_\tau=204$ (Fig.~\ref{fig5}). Within $y^+=10$, the rms fluctuation velocities show improvement by filtering. Also, the location of maximum $u^{''+}_1$ for the $P_f=1$ curve matches the DNS. The modal filtering model over-estimates the fluctuation velocities in the range $0<y^+<40$ compared to DNS results, yet it provides a better estimation of fluctuation velocities than the coarse DNS. Considering fluctuation velocities for $Re_\tau=395$ (Fig.~\ref{fig6}) more points in $y^+=10$ improves the performance of modal filtering in determining $u_2^{''+}$ and $u_3^{''+}$. In Fig.~\ref{fig7}, the modal filter marginally improves the stream-wise rms fluctuation velocity compared to coarse DNS for $Re_\tau=540$. However, the modal filtering model performs better in predicting $u_2^{''+}$ and $u_3^{''+}$.


\begin{figure}[hbt!]
  \begin{center}
    \begin{tikzpicture}
  
      \begin{axis}[
          height=6cm,width=11cm,
          ylabel=$u_1^{''+}$,ylabel style={font=},
          legend style={at={(0.8,0.4)},anchor=north}, 
          x tick label style={anchor=north}, 
          xmin=0,xmax=90,ymin=0,ymax=3
        ]
        \addplot [dashed,color=red,very thick]
        table[x=y_c,y=u_c,col sep=comma] {Re395rms.csv}; 
        \addplot [color=green,very thick]
        table[x=y_l,y=u_l,col sep=comma] {Re395rms.csv}; 
         \addplot [color=black,very thick]
        table[x=y_d,y=u_d,col sep=comma] {Re395rms.csv}; 
        \legend{Coarse DNS,$P_f=1$,DNS \cite{moser1999direct}}
      \end{axis}

    \end{tikzpicture}
    \\
    \begin{tikzpicture}
  
      \begin{axis}[
          height=6cm,width=11cm,
          ylabel=$u_2^{''+}$,ylabel style={font=},
          legend style={at={(0.8,0.9)},anchor=north}, 
          x tick label style={anchor=north}, 
          xmin=0,xmax=90,ymin=0,ymax=1.2
        ]
        \addplot [dashed,color=red,very thick]
        table[x=y_c,y=w_c,col sep=comma] {Re395rms.csv}; 
        \addplot [color=green,very thick]
        table[x=y_l,y=w_l,col sep=comma] {Re395rms.csv}; 
         \addplot [color=black,very thick]
        table[x=y_d,y=v_d,col sep=comma] {Re395rms.csv}; 
      \end{axis}

    \end{tikzpicture}
    \\
    \begin{tikzpicture}
  
      \begin{axis}[
          height=6cm,width=11cm,
          xlabel=$y^+$,ylabel style={font=}, 
          ylabel=$u_3^{''+}$,xlabel style={font=},
          legend style={at={(0.7,0.9)},anchor=north}, 
          x tick label style={anchor=north}, 
          xmin=0,xmax=90,ymin=0,ymax=1.5
        ]
        \addplot [dashed,color=red,very thick]
        table[x=y_c,y=v_c,col sep=comma] {Re395rms.csv}; 
        \addplot [color=green,very thick]
        table[x=y_l,y=v_l,col sep=comma] {Re395rms.csv}; 
         \addplot [color=black,very thick]
        table[x=y_d,y=w_d,col sep=comma] {Re395rms.csv}; 
      \end{axis}

    \end{tikzpicture}
    \caption{Stream-wise, wall-normal, and span-wise rms velocity fluctuations, compared with DNS of Moser et al.  \cite{moser1999direct}, for CHN395 }
    \label{fig6}
  \end{center}
  
\end{figure}
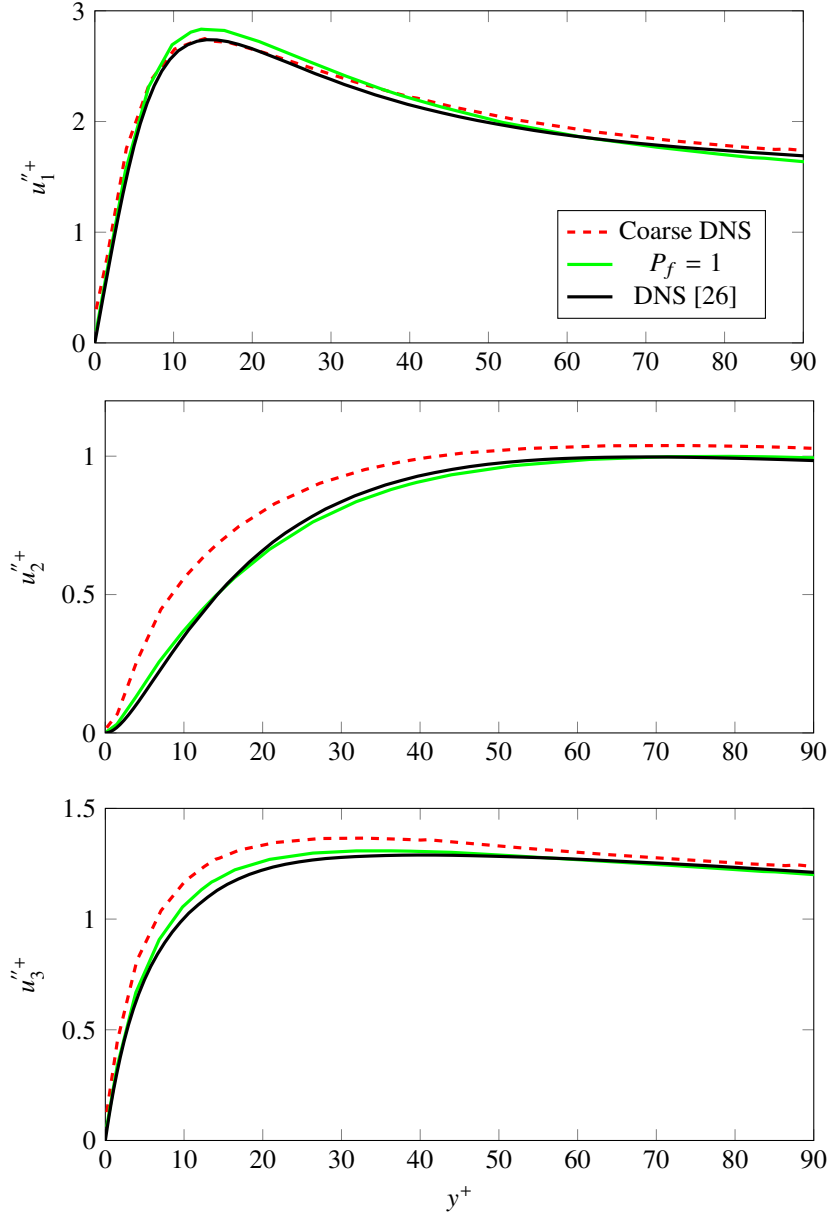
\begin{figure}[hbt!]
  \begin{center}
    \begin{tikzpicture}
  
      \begin{axis}[
          height=6cm,width=11cm,
          ylabel=$u_1^{''+}$,ylabel style={font=},
          legend style={at={(0.8,0.4)},anchor=north}, 
          x tick label style={anchor=north}, 
          xmin=0,xmax=90,ymin=0,ymax=3
        ]
        \addplot [dashed,color=red,very thick]
        table[x=y_c,y=u_c,col sep=comma] {Re540rms.csv}; 
        \addplot [color=green,very thick]
        table[x=y_l,y=u_l,col sep=comma] {Re540rms.csv}; 
         \addplot [color=black,very thick]
        table[x=y_d,y=u_d,col sep=comma] {Re540rms.csv}; 
        \legend{Coarse DNS,$P_f=1$,DNS \cite{lee2015direct}}
      \end{axis}

    \end{tikzpicture}
    \\
    \begin{tikzpicture}
  
      \begin{axis}[
          height=6cm,width=11cm,
          ylabel=$u_2^{''+}$,ylabel style={font=},
          legend style={at={(0.8,0.9)},anchor=north}, 
          x tick label style={anchor=north}, 
          xmin=0,xmax=90,ymin=0,ymax=1.2
        ]
        \addplot [dashed,color=red,very thick]
        table[x=y_c,y=w_c,col sep=comma] {Re540rms.csv}; 
        \addplot [color=green,very thick]
        table[x=y_l,y=w_l,col sep=comma] {Re540rms.csv}; 
         \addplot [color=black,very thick]
        table[x=y_d,y=v_d,col sep=comma] {Re540rms.csv}; 
      \end{axis}

    \end{tikzpicture}
    \\
    \begin{tikzpicture}
  
      \begin{axis}[
          height=6cm,width=11cm,
          xlabel=$y^+$,ylabel style={font=}, 
          ylabel=$u_3^{''+}$,xlabel style={font=},
          legend style={at={(0.7,0.9)},anchor=north}, 
          x tick label style={anchor=north}, 
          xmin=0,xmax=90,ymin=0,ymax=1.5
        ]
        \addplot [dashed,color=red,very thick]
        table[x=y_c,y=v_c,col sep=comma] {Re540rms.csv}; 
        \addplot [color=green,very thick]
        table[x=y_l,y=v_l,col sep=comma] {Re540rms.csv}; 
         \addplot [color=black,very thick]
        table[x=y_d,y=w_d,col sep=comma] {Re540rms.csv}; 
      \end{axis}

    \end{tikzpicture}
    \caption{Stream-wise, wall-normal, and span-wise rms velocity fluctuations, compared with DNS of Lee and Moser  \cite{lee2015direct}, for CHN540 }
    \label{fig7}
  \end{center}
 
\end{figure}
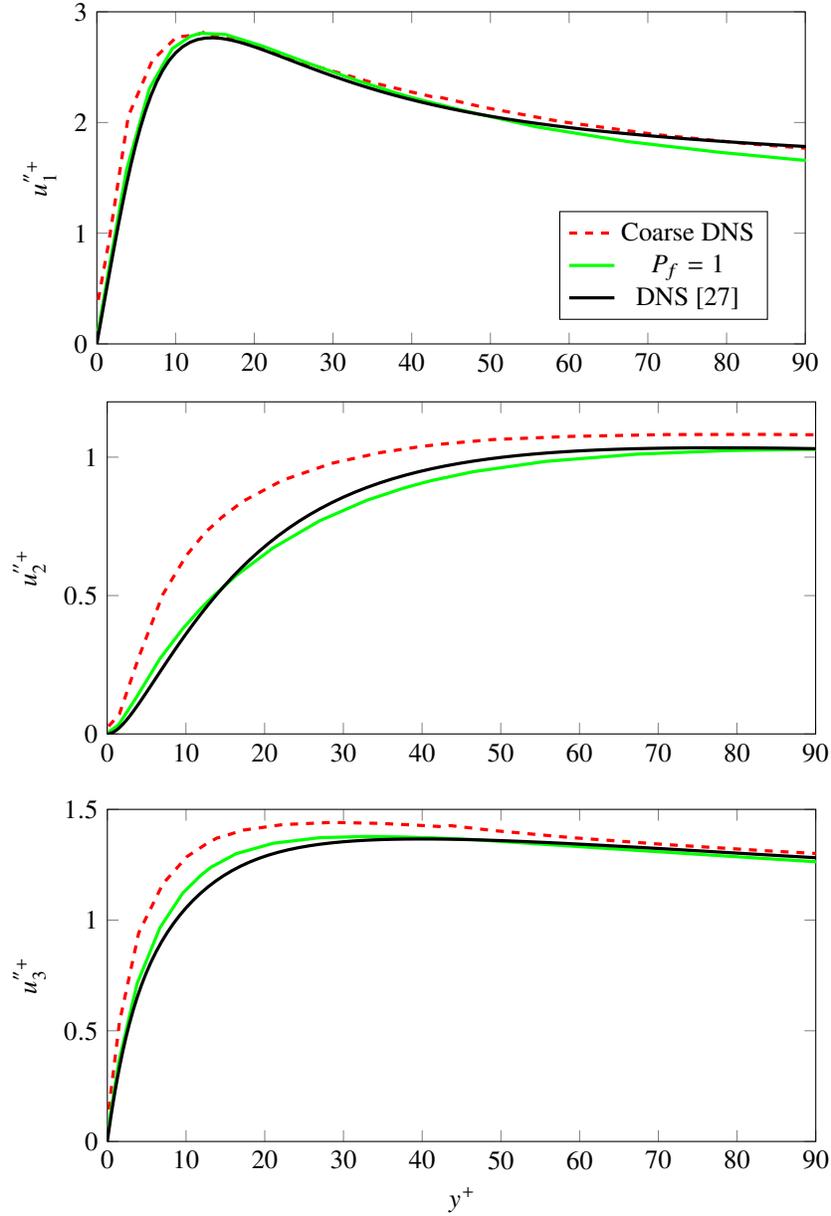
\begin{figure}[hbt!]
  \begin{center}
    \begin{tikzpicture}
  
      \begin{axis}[
          height=6cm,width=11cm,
          ylabel=$TKE$,ylabel style={font=},
          legend style={at={(0.7,0.9)},anchor=north}, 
          x tick label style={anchor=north}, 
          xmin=0,xmax=90,ymin=0,ymax=6
        ]
        \addplot [dashed,color=red,very thick]
        table[x=y_c,y=TKE_c,col sep=comma] {Re204rms.csv}; 
        \addplot [color=green,very thick]
        table[x=y_l,y=TKE_l,col sep=comma] {Re204rms.csv}; 
         \addplot [color=black,very thick]
        table[x=y_d,y=TKE_d,col sep=comma] {Re204rms.csv}; 
        \legend{Coarse DNS,$P_f=1$,{DNS \cite{ghiasi2018near}, CHN200}}
      \end{axis}

    \end{tikzpicture}
    \\
    \begin{tikzpicture}
  
      \begin{axis}[
          height=6cm,width=11cm,
          ylabel=$TKE$,ylabel style={font=},
          legend style={at={(0.7,0.4)},anchor=north}, 
          x tick label style={anchor=north}, 
          xmin=0,xmax=90,ymin=0,ymax=6
        ]
        \addplot [dashed,color=red,very thick]
        table[x=y_c,y=TKE_c,col sep=comma] {Re395rms.csv}; 
        \addplot [color=green,very thick]
        table[x=y_l,y=TKE_l,col sep=comma] {Re395rms.csv}; 
         \addplot [color=black,very thick]
        table[x=y_d,y=TKE_d,col sep=comma] {Re395rms.csv}; 
        \legend{Coarse DNS,$P_f=1$,{DNS \cite{moser1999direct}, CHN395}}
      \end{axis}

    \end{tikzpicture}
    \\
    \begin{tikzpicture}
  
      \begin{axis}[
          height=6cm,width=11cm,
          xlabel=$y^+$,ylabel style={font=}, 
          ylabel=$TKE$,xlabel style={font=},
          legend style={at={(0.7,0.4)},anchor=north},
          x tick label style={anchor=north}, 
          xmin=0,xmax=90,ymin=0,ymax=6
        ]
        \addplot [dashed,color=red,very thick]
        table[x=y_c,y=TKE_c,col sep=comma] {Re540rms.csv}; 
        \addplot [color=green,very thick]
        table[x=y_l,y=TKE_l,col sep=comma] {Re540rms.csv}; 
         \addplot [color=black,very thick]
        table[x=y_d,y=TKE_d,col sep=comma] {Re540rms.csv}; 
        \legend{Coarse DNS,$P_f=1$,{DNS \cite{lee2015direct}, CHN540}}
      \end{axis}

    \end{tikzpicture}
    \caption{Turbulent kinetic energy (TKE) profile in wall normal distance for CHN200, CHN395, and CHN540}
    \label{fig8}
  \end{center}
 
\end{figure}
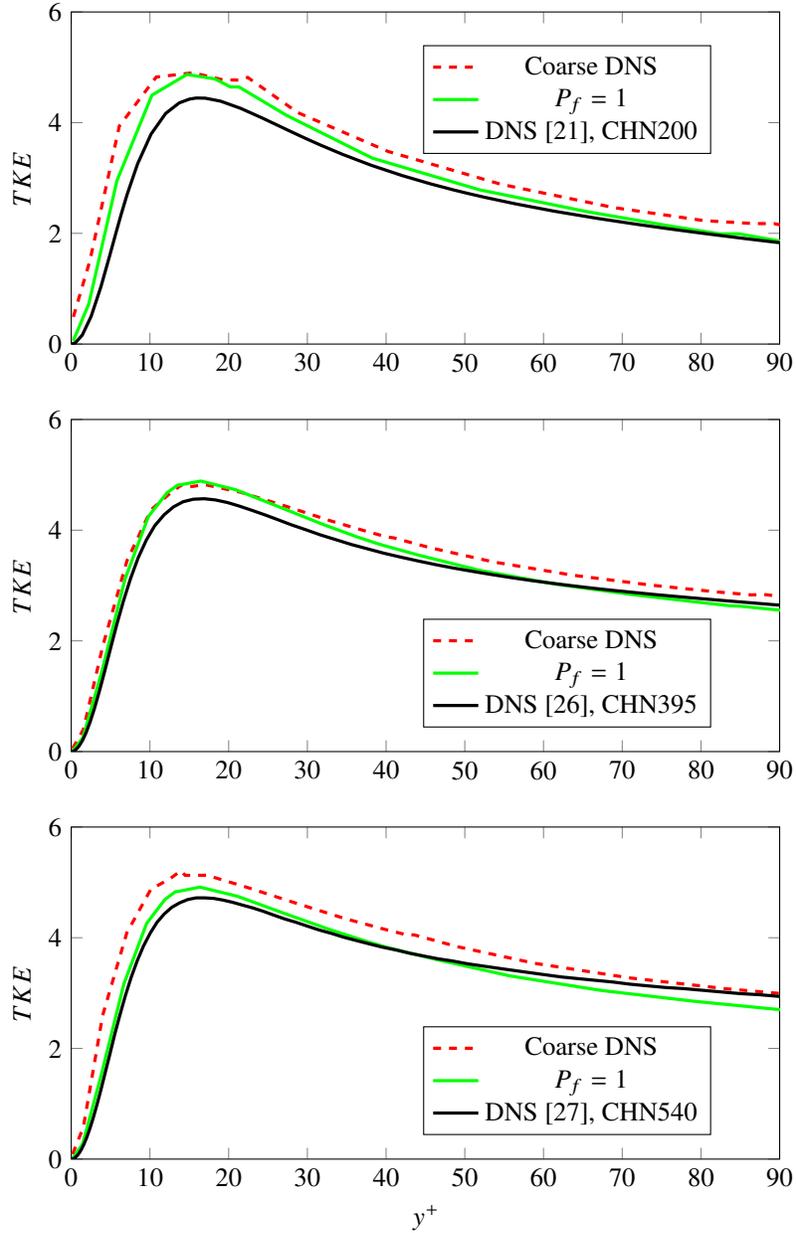

The turbulent kinetic energy, defined as $TKE=0.5({u_1^{''+}}^2+{u_2^{''+}}^2+{u_3^{''+}}^2)$, is plotted for three Reynolds numbers in Fig.~\ref{fig8}. For all the cases, the coarse DNS solution contains more turbulent kinetic energy than DNS. The modal filter removes a portion of the extra energy that is accumulated at the mesh resolution cut-off wavenumber. For $y^+>50$, as the Reynolds number increases, the filter with the strength of $P_f=1$ drains more energy than is required, i.e., the $P_f=1$ curve falls below the DNS curve. In $y^+<50$, the $P_f=1$ overshoots the DNS curve, meaning more energy must be removed in this region. As a result, we can conclude that the filter strength should change with wall-normal distance.

\section{Conclusions}
The explicit modal filtering technique has been employed for the LES of three fully developed turbulent channel flows, with $Re_\tau=204, 395$ and $544$. In modal filtering LES, the governing equations of the compressible flow are solved for a coarse grid resolution using DSEM with no additional SGS model. The best configuration of the strength and the frequency of filtering is chosen for the simulations. The LES model can predict mean profile accurately for all the friction Reynolds numbers. It also estimates the rms statistics quite accurately for span-wise and wall-normal directions, whereas for the stream-wise fluctuation velocity, it produces a slight overshoot at the location of maximum fluctuation intensity near the wall. The mode removal procedure locally drains turbulent kinetic energy on each element. The LES results show that the kinetic energy level is lower than in DNS away from the wall, which leads to the conclusion that the modal filtering method requires regularization in terms of the wall distance. The LES model is designed to be computationally inexpensive since it does not require the additional calculations of closure terms in the filtered Navier-Stokes equations. Therefore, it is essential to examine the method performance in different turbulent flows such as wall-bounded, plane mixing layer, and the plane wake flows. Moreover, the explicit modal filtering is a perfect choice for the LES of complex geometry flows since it is implemented locally for each element and it has negligible computational overload.

\FloatBarrier 
\section*{Acknowledgments}
The authors acknowledge the Advanced Cyberinfrastructure for Education and Research (ACER) group at The University of Illinois at Chicago for providing high performance computing (HPC) resources that have contributed to the results reported in this paper. We would like to also thank the administration of XSEDE’s supercomputer, Comet, who provided us with computational resources and technical support under the grant TG-CTS180014.
\newpage
\bibliography{sample}

\begin{thebibliography}{28}
\newcommand{\enquote}[1]{``#1''}
\providecommand{\natexlab}[1]{#1}
\providecommand{\url}[1]{\texttt{#1}}
\providecommand{\urlprefix}{URL }
\expandafter\ifx\csname urlstyle\endcsname\relax
  \providecommand{\doi}[1]{doi:\discretionary{}{}{}#1}\else
  \providecommand{\doi}{doi:\discretionary{}{}{}\begingroup
  \urlstyle{rm}\Url}\fi

\bibitem[{Li et~al.(2019)Li, Komperda, Ghiasi, Peyvan, and
  Mashayek}]{li2019compressibility}
Li, D., Komperda, J., Ghiasi, Z., Peyvan, A., and Mashayek, F.,
  \enquote{Compressibility effects on the transition to turbulence in a
  spatially developing plane free shear layer,} \emph{Theoretical and
  Computational Fluid Dynamics}, Vol.~33, No.~6, 2019, pp. 577--602.

\bibitem[{Abtahi et~al.(2019)Abtahi, Rosti, Mirbod, and
  Brandt}]{abtahi2019porous}
Abtahi, S., Rosti, M., Mirbod, P., and Brandt, L., \enquote{Porous walls impact
  on suspension flows,} \emph{APS}, 2019, pp. M04--039.

\bibitem[{Ghiasi et~al.(2019)Ghiasi, Komperda, Li, Peyvan, Nicholls, and
  Mashayek}]{ghiasi2019modal}
Ghiasi, Z., Komperda, J., Li, D., Peyvan, A., Nicholls, D., and Mashayek, F.,
  \enquote{Modal explicit filtering for large eddy simulation in discontinuous
  spectral element method,} \emph{Journal of Computational Physics: X}, Vol.~3,
  2019, p. 100024.

\bibitem[{Peyvan and Benisi(2016)}]{peyvan2016axial}
Peyvan, A., and Benisi, A., \enquote{Axial-Flow Compressor Performance
  Prediction in Design and Off-Design Conditions through 1-D and 3-D Modeling
  and Experimental Study.} \emph{Journal of Applied Fluid Mechanics}, Vol.~9,
  No.~5, 2016.

\bibitem[{Pope(2001)}]{pope2001turbulent}
Pope, S.~B., \emph{Turbulent flows}, IOP Publishing, 2001.

\bibitem[{Smagorinsky(1963)}]{smagorinsky1963general}
Smagorinsky, J., \enquote{General circulation experiments with the primitive
  equations: I. The basic experiment,} \emph{Monthly weather review}, Vol.~91,
  No.~3, 1963, pp. 99--164.

\bibitem[{Bardina et~al.(1980)Bardina, Ferziger, and
  Reynolds}]{bardina1980improved}
Bardina, J., Ferziger, J., and Reynolds, W., \enquote{Improved subgrid-scale
  models for large-eddy simulation,} \emph{13th Fluid and Plasma Dynamics
  Conference}, 1980, p. 1357.

\bibitem[{Germano et~al.(1991)Germano, Piomelli, Moin, and
  Cabot}]{germano1991dynamic}
Germano, M., Piomelli, U., Moin, P., and Cabot, W.~H., \enquote{A dynamic
  subgrid-scale eddy viscosity model,} \emph{Physics of Fluids A: Fluid
  Dynamics}, Vol.~3, No.~7, 1991, pp. 1760--1765.

\bibitem[{Moin et~al.(1991)Moin, Squires, Cabot, and Lee}]{moin1991dynamic}
Moin, P., Squires, K., Cabot, W., and Lee, S., \enquote{A dynamic subgrid-scale
  model for compressible turbulence and scalar transport,} \emph{Physics of
  Fluids A: Fluid Dynamics}, Vol.~3, No.~11, 1991, pp. 2746--2757.

\bibitem[{Erlebacher et~al.(1992)Erlebacher, Hussaini, Speziale, and
  Zang}]{erlebacher1992toward}
Erlebacher, G., Hussaini, M.~Y., Speziale, C.~G., and Zang, T.~A.,
  \enquote{Toward the large-eddy simulation of compressible turbulent flows,}
  \emph{Journal of Fluid Mechanics}, Vol. 238, 1992, pp. 155--185.

\bibitem[{Lenormand et~al.(2000)Lenormand, Sagaut, and
  Ta~Phuoc}]{lenormand2000large}
Lenormand, E., Sagaut, P., and Ta~Phuoc, L., \enquote{Large eddy simulation of
  subsonic and supersonic channel flow at moderate Reynolds number,}
  \emph{International Journal for Numerical Methods in Fluids}, Vol.~32, No.~4,
  2000, pp. 369--406.

\bibitem[{Martin et~al.(2000)Martin, Piomelli, and Candler}]{martin2000subgrid}
Martin, M.~P., Piomelli, U., and Candler, G.~V., \enquote{Subgrid-scale models
  for compressible large-eddy simulations,} \emph{Theoretical and Computational
  Fluid Dynamics}, Vol.~13, No.~5, 2000, pp. 361--376.

\bibitem[{Lodato et~al.(2013)Lodato, Castonguay, and
  Jameson}]{lodato2013discrete}
Lodato, G., Castonguay, P., and Jameson, A., \enquote{Discrete filter operators
  for large-eddy simulation using high-order spectral difference methods,}
  \emph{International Journal for Numerical Methods in Fluids}, Vol.~72, No.~2,
  2013, pp. 231--258.

\bibitem[{Vasilyev et~al.(1998)Vasilyev, Lund, and Moin}]{vasilyev1998general}
Vasilyev, O.~V., Lund, T.~S., and Moin, P., \enquote{A general class of
  commutative filters for LES in complex geometries,} \emph{Journal of
  Computational Physics}, Vol. 146, No.~1, 1998, pp. 82--104.

\bibitem[{Grinstein and Fureby(2002)}]{grinstein2002recent}
Grinstein, F.~F., and Fureby, C., \enquote{Recent progress on MILES for high
  Reynolds number flows,} \emph{Journal of Fluids Engineering}, Vol. 124,
  No.~4, 2002, pp. 848--861.

\bibitem[{Gassner and Beck(2013)}]{gassner2013accuracy}
Gassner, G.~J., and Beck, A.~D., \enquote{On the accuracy of high-order
  discretizations for underresolved turbulence simulations,} \emph{Theoretical
  and Computational Fluid Dynamics}, Vol.~27, No. 3-4, 2013, pp. 221--237.

\bibitem[{Flad et~al.(2016)Flad, Beck, and Munz}]{flad2016simulation}
Flad, D., Beck, A., and Munz, C.-D., \enquote{Simulation of underresolved
  turbulent flows by adaptive filtering using the high order discontinuous
  Galerkin spectral element method,} \emph{Journal of Computational Physics},
  Vol. 313, 2016, pp. 1--12.

\bibitem[{Winters et~al.(2018)Winters, Moura, Mengaldo, Gassner, Walch, Peiro,
  and Sherwin}]{winters2018comparative}
Winters, A.~R., Moura, R.~C., Mengaldo, G., Gassner, G.~J., Walch, S., Peiro,
  J., and Sherwin, S.~J., \enquote{A comparative study on polynomial dealiasing
  and split form discontinuous Galerkin schemes for under-resolved turbulence
  computations,} \emph{Journal of Computational Physics}, 2018.

\bibitem[{Komperda et~al.(2020{\natexlab{a}})Komperda, Ghiasi, Li, Peyvan,
  Jaberi, and Mashayek}]{komperda2020hybrid}
Komperda, J., Ghiasi, Z., Li, D., Peyvan, A., Jaberi, F., and Mashayek, F.,
  \enquote{A hybrid discontinuous spectral element method and filtered mass
  density function solver for turbulent reacting flows,} \emph{Numerical Heat
  Transfer, Part B: Fundamentals}, Vol.~78, No.~1, 2020{\natexlab{a}}, pp.
  1--29.

\bibitem[{Komperda et~al.(2020{\natexlab{b}})Komperda, Li, Peyvan, and
  Mashayek}]{komperda2020filtered}
Komperda, J., Li, D., Peyvan, A., and Mashayek, F., \enquote{Filtered Density
  Function for Shocked Compressible Flows on Unstructured Spectral Element
  Grids,} \emph{AIAA Scitech 2020 Forum}, 2020{\natexlab{b}}, p. 1789.

\bibitem[{Ghiasi et~al.(2018)Ghiasi, Li, Komperda, and
  Mashayek}]{ghiasi2018near}
Ghiasi, Z., Li, D., Komperda, J., and Mashayek, F., \enquote{Near-wall
  resolution requirement for direct numerical simulation of turbulent flow
  using multidomain Chebyshev grid,} \emph{International Journal of Heat and
  Mass Transfer}, Vol. 126, 2018, pp. 746--760.

\bibitem[{Kopriva and Kolias(1996)}]{kopriva1996conservative}
Kopriva, D.~A., and Kolias, J.~H., \enquote{A conservative staggered-grid
  Chebyshev multidomain method for compressible flows,} \emph{Journal of
  Computational Physics}, Vol. 125, No.~1, 1996, pp. 244--261.

\bibitem[{Toro(2013)}]{toro2013riemann}
Toro, E.~F., \emph{Riemann solvers and numerical methods for fluid dynamics: a
  practical introduction}, Springer Science \& Business Media, 2013.

\bibitem[{Jacobs(2003)}]{jacobs2003numerical}
Jacobs, G., \enquote{Numerical Simulation of Two-phase Turbulent Compressible
  Flows with a Multidomain Spectral Method,} Ph.D. thesis, 2003.

\bibitem[{Kopriva(2009)}]{kopriva2009implementing}
Kopriva, D.~A., \emph{Implementing spectral methods for partial differential
  equations: Algorithms for scientists and engineers}, Springer Science \&
  Business Media, 2009.

\bibitem[{Moser et~al.(1999)Moser, Kim, and Mansour}]{moser1999direct}
Moser, R.~D., Kim, J., and Mansour, N.~N., \enquote{Direct numerical simulation
  of turbulent channel flow up to Re $\tau$= 590,} \emph{Physics of fluids},
  Vol.~11, No.~4, 1999, pp. 943--945.

\bibitem[{Lee and Moser(2015)}]{lee2015direct}
Lee, M., and Moser, R.~D., \enquote{Direct numerical simulation of turbulent
  channel flow up to $Re_\tau=5200$,} \emph{Journal of Fluid Mechanics}, Vol.
  774, 2015, pp. 395--415.

\bibitem[{Choi and Moin(2012)}]{choi2012grid}
Choi, H., and Moin, P., \enquote{Grid-point requirements for large eddy
  simulation: Chapman’s estimates revisited,} \emph{Physics of Fluids},
  Vol.~24, No.~1, 2012, p. 011702.

\end{thebibliography}

\end{document}